# Optical generation of non-diffracting beams via photorefractive holography

Tárcio A. Vieira,[1] Rafael A. B. Suarez,[1] Marcos R. R. Gesualdi,[1,*] Michel Zamboni-Rached[2]

[1]*Universidade Federal do ABC, Av. dos Estados 5001, CEP 09210-580, Santo André, SP, Brazil*
[2]*Departamento de Comunicações, Faculdade de Engenharia Elétrica, Universidade Estadual de Campinas, Campinas, SP, Brazil*
*\*Corresponding author: [marcos.gesualdi@ufabc.edu.br](mailto:marcos.gesualdi@ufabc.edu.br)*

**This work presents, for the first time the optical generation of non-diffracting beams via photorefractive holography. Optical generation of non-diffracting beams using conventional optics components is difficult and, in some instances, unfeasible, as it is wave fields given by superposition of non-diffracting beams. It is known that computer generated holograms and spatial light modulators (SLMs) successfully generate such beams. With photorefractive holography technique, the hologram of a non-diffracting beam is constructed (recorded) and reconstructed (reading) optically in a nonlinear photorefractive medium. The experimental realization of a non-diffracting beam was made in a photorefractive holography setup using a photorefractive $Bi_{12}SiO_{20}$ (BSO) crystal as the holographic recording medium, where the non-diffracting beams, the Bessel beam arrays and superposition of co-propagating Bessel beams (Frozen Waves) were obtained experimentally. The experimental results are in agreement with the theoretically predicted results, presenting excellent prospects for implementation of this technique for dynamical systems at applications in optics and photonics.**

## 1. INTRODUCTION

Localized waves or non-diffracting or diffraction-resistant waves are wave packets that remain unchanged during propagation [1]. The types of non-diffracting beams include Bessel beams, Mathieu beams, Parabolic beams and Airy beams[2-5]; as well as the superposition of these beams which produces special optical beams such as Frozen Waves, co-propagating superposition of Bessel beams [10,11]. We can visualize these optical beams various fields of applications in optical metrology, optical alignment systems over long distances and mechanical arrangements; optical tweezers for handling micro structures; in nonlinear optics, optical coherence tomography; imaging systems, optical communications in free space; among others [5,6,7,8,9].

Experimental generation of non-diffracting beams using conventional diffractive optical components presents several difficulties, and in some cases is not feasible, as co-propagating beam overlap of this type.

Holography has been presented as a powerful tool for generating special optical beams and other applications in the field of optics and photonics [12-23].

Thus, the computational holography technique [12-14] with the use of numeric holograms and spatial light modulators to efficiently reproduce such beams is successful [14-16]. In this case, the construction of the non-diffracting beam hologram is done numerically by computer generated hologram (CGH) and is reconstructed optically with its implementation in a spatial light modulator (SLM), which is a suitable optical arrangement that enables its experimental generation [14-16].

On the other hand, photorefractive holography has been presented as a promising technique for dynamic holographic processes and holographic interferometry methods to analyze surfaces and optical wavefronts [17-23]. This is based on the photorefractive effect, consisting in modulation of the refractive index via photoinduction of charge carriers and linear electro-optic effect in some semiconductors crystals with special features, so-called photorefractive crystals ($LiNbO_3$, SBN, KBT, BaTiO3, $Bi_{12}SiO_{20}$, among others) [17]. Due to the fact that it is a process that occurs the electronic level in semiconductor crystals with nonlinear optical properties, the holographic gratings feature high resolution and short response time, making it possible to act as holographic recording media. This not require chemical or computational processing for reconstruction of the holographic image and present indefinite reusability [17-23].

In this work we use the photorefractive holographic technique, where in the hologram of the non-diffracting beam is constructed (recorded) and reconstructed (reading) optically in a photorefractive non-linear medium by refractive index modulation generated via a second order non linear effect (Pockels effect). The experimental realization of non-diffracting beam was made in a photorefractive holography setup using a photorefractive sillenite crystal type $Bi_{12}SiO_{20}$ (BSO), as the holographic recording medium. Thus, Bessel beam, Mathieus beam and Parabolic beam; as well as the non-diffracting beams arrays and superposition of co-propagating Bessel beams (Frozen waves) were generated experimentally with excellent agreement with the theoretical predictions and the obtained results using computational holography technique. These beams obtained presents excellent prospects for implementation in dynamic systems for scientific and technological applications.

## 2. THEORETICAL BACKGROUND

### A. Non-diffracting beams

Non-diffracting waves are wave packets that remain unchanged during propagation [1]. The Helmholtz equation, $[\nabla^2 + k_0^2]\Psi = 0$, describes the propagation of light considering the effects of diffraction and scattering. In 1987, J. Durnin published an important work showing that the Helmholtz equation has a family of solutions that propagate diffraction free [2]. He obtained these solutions to the Helmholtz equation using the cylindrical coordinate system, now known as Bessel beams. The general expression of a Bessel beam $\nu$ order that propagates in the positive direction of the $z$ axis is:

$$\Psi(\rho,\phi,z) = J_\nu(k_T \rho)\, e^{ik_z z} e^{i\nu\phi} \qquad (1)$$

Where $J_\nu$ is the Bessel function of $\nu$ order; $k_T$ is the transverse wave number, which defines the lateral width of the beam. Taking ($\rho$) intensity, we see that the cross-section follows a Bessel function of 0 order and independent of $z$

$$I(\rho) = |J_\nu(k_T \rho)|^2 \qquad (2)$$

Other types of non-diffracting beams can be created with the same treatment to obtain solutions of the Helmholtz equation: using an elliptical-cylindrical coordinate system, we get the so-called Mathieus beams [3]; and, for a parabolic-cylindrical coordinate system we obtain a non-diffracting beam with parabolic transversal shape [4]. In equations (3-4) we have an even solution for the Helmholtz equation in the respective coordinates:

$$\Psi_e(\eta,\xi,z) = ce_m(\eta,q) Je_m(\xi,q) \qquad (3)$$

$$\Psi_e(\mu,\nu',a) = \frac{1}{\pi\sqrt{2}} |\Gamma_1|^2 P_e(\sigma\nu';a) P_e(\sigma\mu';-a) \qquad (4)$$

Also, the non-diffracting beam arrays structure has become extremely promising in applications such as optical tweezers and optical communications [5].

On the other hand, the superposition of co-propagating Bessel beams generating a special types of non-diffracting beam was called Frozen Wave [6-7]. Basically, the idea is to obtain non-diffracting beams whose desired longitudinal intensity pattern, $|F(z)|^2$, in the interval $0 \leq z \leq L$, can be chosen a priori. To obtain the desired beam we consider the following solution, given by a superposition of 2N + 1 co-propagating and equal frequency Bessel beams of order $\nu$:

$$\Psi(\rho,\phi,z,t) = e^{-i\omega t} \sum_{n=-N}^{N} A_n J_\nu(k_{\rho n}\rho) e^{ik_{zn}z} e^{i\nu\phi} \quad (5)$$

whereas $k_{\rho n}^2 = \omega^2/c^2 - k_{zn}^2$ and with the choice $k_{zn} = Q + 2\pi n/L$ the $Q$ parameter is a constant obeying $0 \leq Q \pm (2\pi/L)N \leq \omega/c$.

In Eq. (1), the coefficients $A_n$ are given by

$$A_n = \frac{1}{L}\int_0^L F(z) e^{-i\frac{2\pi}{L}nz} dz \quad (6)$$

whereas $|F(z)|^2$ is the desired longitudinal intensity pattern in the interval $0 \leq z \leq L$. This longitudinal intensity pattern can be concentrated (as we wish) over; along the propagation axis ($\rho = 0$) taking $\nu = 0$ in Eq. (5), i.e., we deal with zero order Bessel beam superposition, or a cylindrical surface. In this case we deal with ν > 0 values [7].

## B. Photorefractive holography

In photorefractive holography using photorefractive crystal (PRC) $Bi_{12}SiO_{20}$ as the recording medium [17-23], the holographic recording occurs by a refractive index modulation, via photorefractive effect in diffusive regimen. The refractive index modulation of the hologram is written as $\Delta n = n_0^3 r_{41} E_{sc}/2$, where $n_0$ is the refractive index, $r_{41}$ is the linear electro-optic coefficient of the sillenite crystal and $E_{sc}$ is the electric field generated by the redistributed space charges in the crystal. And the holographic reconstruction of the object wavefront occur in quasi real-time, where the optical reconstruction of the holographic image is in diffracted wave. If λ is the recording wavelength, ρ is the crystal rotator power, L is the crystal thickness, m is the modulation of the incident interference pattern and $2\theta$ is the angle between the interfering beams, the diffraction efficiency of a hologram grating recorded in a PR crystal is given by [17-23]

$$\eta = \left(\frac{\pi \Delta n}{\lambda \cos\theta} \frac{\sin \rho L}{\rho L}\right)^2 m^2 \quad (7)$$

The intensity $I_0$ at a point $(x, y)$ resulting from the superposition of the diffracted ($I_{0,D}$) and the transmitted ($I_{0,T}$) intensities is given by

$$I_0(x,y) = I_{0,T}(x,y) + I_{0,D}(x,y)\left[1 - e^{(-t/\tau)}\right]^2 \quad (8)$$

whereas, τ is the hologram response (writing or erasure) time. The holographic reconstruction of the object wave, $I_{0,D}(x,y)$, is written as

$$I_{0,D}(x,y) = I_{0,O}(x,y)\eta + I_{0,R}(x,y)[1-\eta] + 2g\Upsilon\cos\Delta\Phi \quad (9)$$

whereas $I_{0,O}(x,y)$ and $I_{0,R}(x,y)$ are object and reference beam intensities, respectively, g is a parameter of the polarization coupling of the beams, $2g\Upsilon cos\Delta\Phi$ is the interference term and $\Delta\Phi$ is the phase shift on the object beam [17-23].

## 3. EXPERIMENTS AND RESULTS

In our investigation, we optimize an experimental setup for photorefractive holography, shown in Figure 1, where a silenite $Bi_{12}SiO_{20}$ (BSO) crystal was the holographic recording medium (with dimensions 10x10x3mm and transverse electro-optical configuration), and a laser Argon (λ = 514,5nm and 1Watt output power at that wavelength the BSO crystal is more sensitive [17]). The reference beam and object incident on the crystal surface at an angle of 45 degrees between them, so we optimize the diffraction efficiency [14-19]. We have knowledge of this method, as we develop and use an experimental apparatus for photorefractive holography to analyze surfaces, as can be seen in references [18-23].

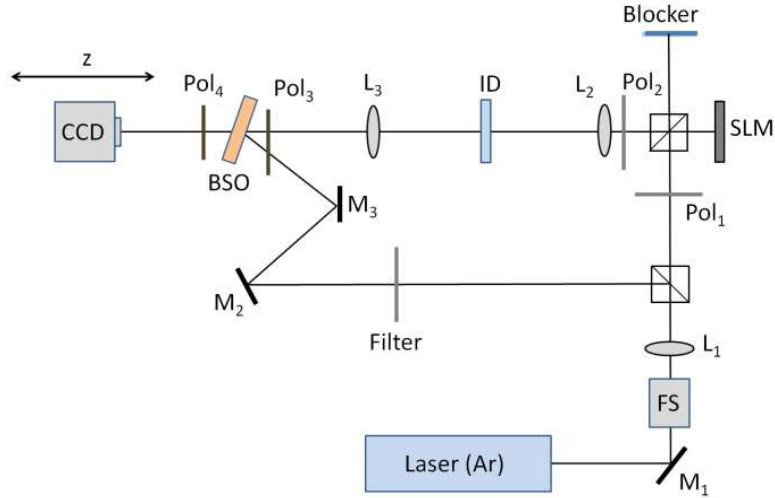

Fig. 1. Photorefractive holographic setup for optical generation of nondiffracting beam, where M's are mirrors, BS is a beam-splitter, FS is the spatial filter, L's are lenses, Pol's are polarizers, ID is the circular mask, BSO is the photorefractive crystal, Filter is neutral density filter and CCD camera for image acquisition.

One of setup arms have SLM to generate a non-diffracting beam which will be recordedin BSO crystal.
As described previously, the photorefractive medium has a characteristic formation time of the spatial-charge field generated by an interference pattern. As a result, the intensity diffracted by the hologram increases exponentially with time until the holographic grating reaches its maximum diffraction efficiency under these conditions. This characteristic time is also related to the holographic grating erasure, in which the uniform intensity of the read beam erases the holographic grating, that decreases exponentially the intensity of the diffracted beam, eq. (8) and (9).

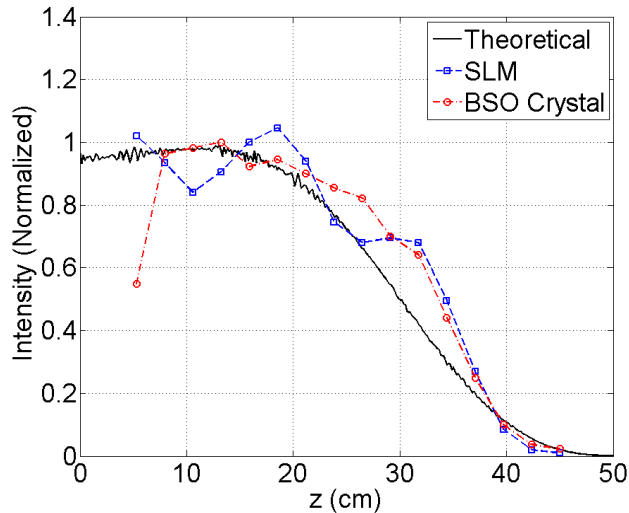

Fig. 2. Comparison between the longitudinal intensity pattern in $\rho = 0$ for a Bessel beam predicted theorically, generated by the SLM and reconstructed via photorefractive holography.

## A. Generating a single non-diffracting beam via photorefractive holography

We apply this technique to optically generate a non-diffracting beam. Initially, the SLM was configured to optically generate a non-diffracting beam (object beam); thus, the object and reference beams focus in the crystal (recording process of the photorefractive hologram), with a saturation time of the holographic grating of approximately 2.8 minutes. Then, the object beam is blocked and only the reference beam focus on the crystal (reading process of the photorefractive hologram), which is diffracted by holographic grating and reconstructing the nondiffracting beam. We captured the transverse patterns of the beam by moving the CCD along the axis of propagation. The images were captured at 2 second intervals from the beginning of the reading process, and spatially displaced by 2.5 cm.

**First case:** Zero order Bessel beam with intensity $J_0^2(k_\rho\rho)$, where $k_\rho = 3.8 \times 10^4 m^{-1}$ and hologram aperture radius R = 1.1mm [14-16], we have a maximum propagation of 40cm. Figure 2 shows the intensity pattern in $\rho = 0$ (on-axis), predicted by theoretical field encoded in numerical hologram (CGH), optical reconstruction by SLM and reconstruction of the hologram recorded in photorefractive medium, Figure 3. We note in Figure 2 that the results are in agreement with the theoretically expected, which was made an apodization using a Gaussian filter in the hologram (CGH) of the Bessel beam; as well as the good results transverse patterns of Bessel beams in Figure 3 optically reconstructed via photorefractive holography are presented.

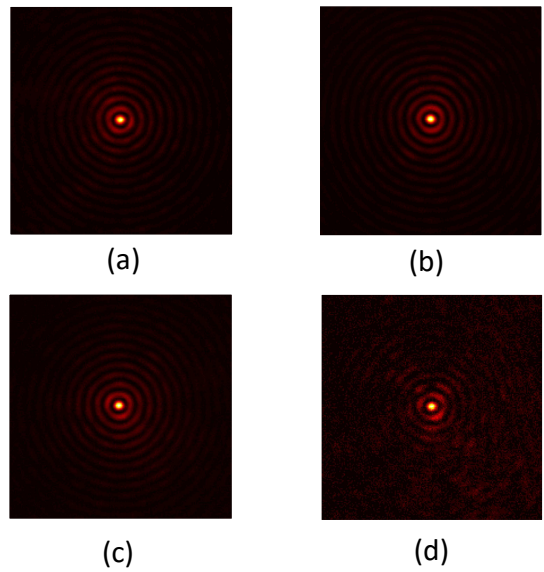

Fig. 3. Longitudinal intensity pattern in $\rho = 0$ for a Bessel beam reconstructed via holography photorefractive with transverse pattern (b) 11 cm, (c) 21 cm, (d) 32cm and (e) 39cm.

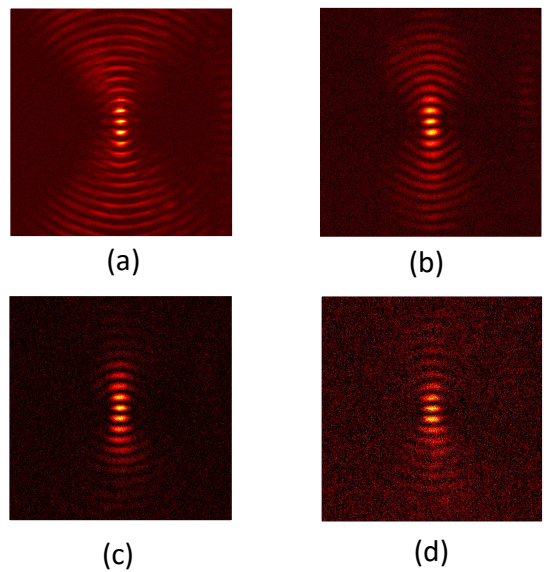

Fig. 4. Longitudinal intensity pattern in $\rho = 0$ for a Mathieu beam reconstructed via holography photorefractive with transverse pattern (b) 10cm, (c) 24cm, (d) 39cm and (e) 51cm..

**Second case:** Similarly to the first case, Mathieu beams were used for experimental validation of the method. Using the equation (3) for the fields to be implemented in numerical hologram and then reconstructed by the SLM, with q = 22 and R = 1.2mm [3]. The results of reconstruction of photorefractive holograms are shown in Figure 4, where we have intensity patterns on the axis of propagation.

**Third case:** Similarly to the previous cases, Parabolic beams were reproduced using the equation (4) to the fields to be implemented in numerical hologram and then reconstructed by the SLM, with a = -5 and R = 0.6mm [4]. The results of photorefractive reconstruction of holograms are shown in Figure 5, where we have intensity patterns on the axis of propagation.

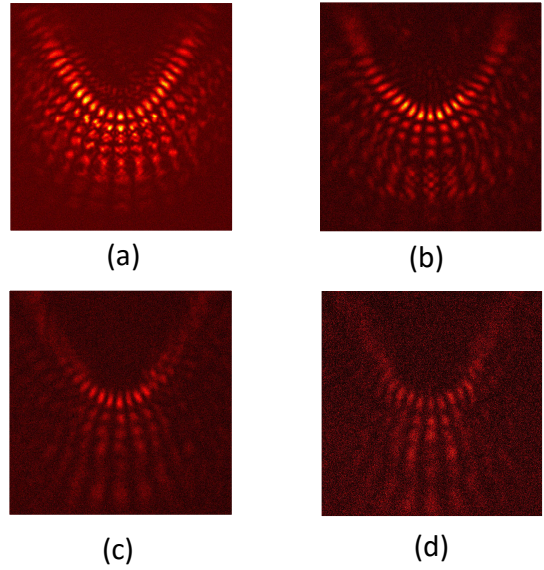

(a)  (b)

(c)  (d)

Fig. 5. Longitudinal intensity pattern in (a) x = 0 e y = 0; 3mm for a Parabolic: reconstructed via holography photorefractive with transverse pattern (b) 11cm, (c) 18cm, (d) 29cm e (e) 40cm.

### B. Generation of non-diffracting beams arrays via photorefractive holography

The process of obtaining numerical fields and non-diffracting beam holograms allows us to generate arrays or a certain spatial distribution of N non-diffracting beams. Arrays of non-diffracting beams are interesting for possible applications in optical tweezers, optical communications for a system with multiple communication channels and for generating solitons in a non linear crystal [24, 25].

From the fields of these different non-diffracting beams $\Psi(x, y, z)$, we consider the origin of the displaced coordinate system on the plane transverse to the direction of propagation $(x\pm\delta x; y\pm\delta y)$. These displaced fields are summed to form a total field $\Psi_{tot} = \sum \psi(x\pm\delta x; y\pm\delta y)$ resulting in an interference pattern and from this, generate a CGH for implementation and reconstruction using an SLM.

Consider a beam composition defined by four zero-order Bessel beams, we have

$$\psi_1(x+\delta x, y+\delta y, z=0) = J_0\left(k_\rho \sqrt{(x+\delta x)^2 + (y+\delta y)^2}\right)$$
$$\psi_2(x-\delta x, y+\delta y, z=0) = J_0\left(k_\rho \sqrt{(x-\delta x)^2 + (y+\delta y)^2}\right)$$
$$\psi_3(x+\delta x, y-\delta y, z=0) = J_0\left(k_\rho \sqrt{(x+\delta x)^2 + (y-\delta y)^2}\right) \quad (10)$$
$$\psi_4(x-\delta x, y-\delta y, z=0) = J_0\left(k_\rho \sqrt{(x-\delta x)^2 + (y-\delta y)^2}\right)$$

And, using the same parameters as the example in Figure 2, but spatially displaced from the center of coordinates ($\delta x$; $\delta y$) $\delta x = \delta y = 1$mm, we have the non-diffracting beam array formed by four zero-order Bessel beams, Figure 6.

Based on the record-reading process of photorefractive holograms, we can register multiple holograms in the crystal over time, and reconstruct them in a single reading process. In this case, different holograms are recorded under the same conditions. However, an inherent problem with photorefractive materials such as BSO is the erasure of recorded holograms while new holograms are recorded by varying the diffraction efficiency and consequently result in non-uniform intensity beams. One solution is to record holograms in shorter time intervals so that the diffraction efficiency of the new holograms accompanies the decay of the first holograms. This approach is called sequential generation and was proposed by Burke [26], who obtained a sequence of 18 holograms recorded with the constant diffraction efficiency. Thus, the following are our results for these cases.

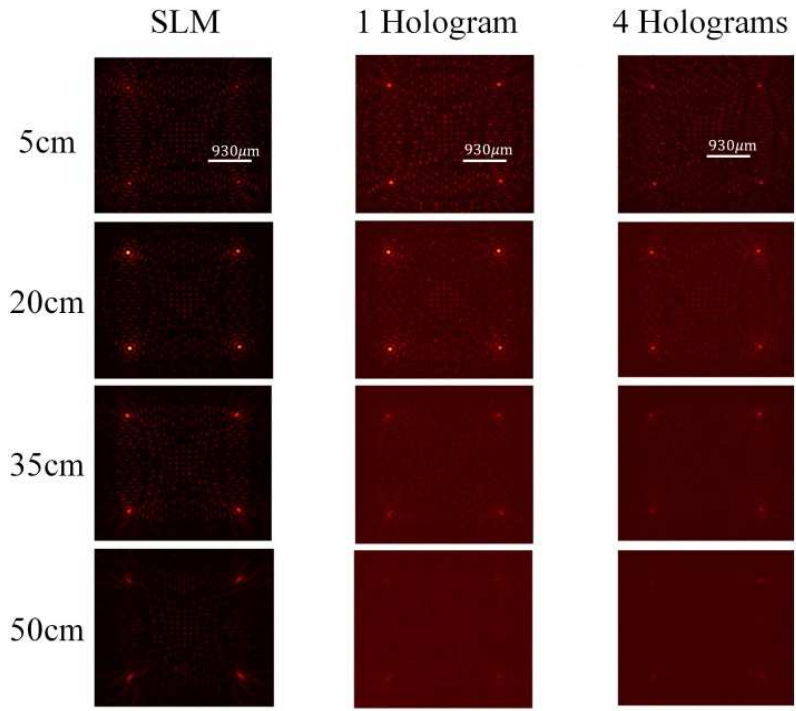

Fig. 6. Optical reconstruction of a matrix with four zero-order Bessel beams using the SLM, with hologram 1 and 4 holograms using photorefractive holography in different positions along the propagation in z = 5cm, 20cm, 35cm, 50cm.

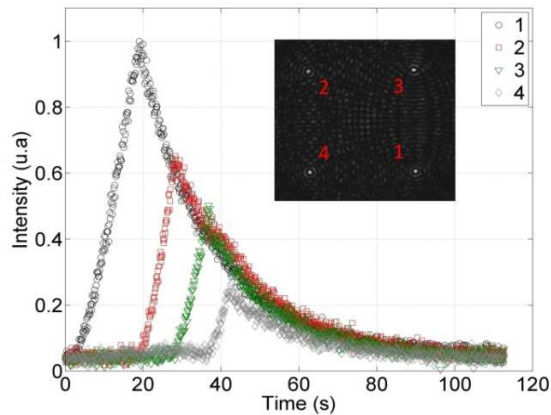

Fig. 7. Sequential recording process of holograms for $t_1$ = 19s, $t_2$ = 11,5s, $t_3$ = 9s, $t_4$ = 6,5s.

In Figure 6, we shown the transverse patterns of the beams along the propagation into the optical reconstruction of a matrix with 4 zero-order Bessel beams using the SLM. With one hologram and 4 holograms using photorefractive holography in different positions along the propagation in z = 5cm, 20cm, 35cm, 50cm.
Using the setup of Figure 1, sequentially recorded on the crystal four beams spatially displaced, keeping the crystal above each Bessel beam for a given time interval $t_i$.

We use parameters such as intensity of reference and object beams, and polarizations identical to those cited above for photorefractive holography in real time. Figure 7 shows a graph with the evolution of the diffracted intensity by holographic grating according to the recording of each hologram. In this particular case, each recorded beam process erased the previously written, due laser incidence on the refractive index grating, so we need that the decay time of the four beams are in synchrony for that in the matrix reconstruction process the intensity of the beam decays with the same rate.

**C. Generation of Frozen waves via photorefractive holography**

We present the following as the most interesting cases from this work, where the photorefractive holography was used to generate Frozen Waves [6,7,11,12]. Considering the possibility of sequential generation of holograms using photorefractive holography, we recorded the superposition of zero-order Bessel beams co-propagating in photorefractive crystal to obtain a longitudinal pattern of specific intensity. The FWs by equations (5) and (6), and using

$$F(z) = \begin{cases} 1, & l_1 \leq z \leq l_2 \\ 0, & elsewhere \end{cases} \quad (11)$$

where $l_1$ = 10cm e $l_2$ = 22cm and L = 40cm.

For Q = 0,99999 k, we have $N_{max}$ = 7, however we will consider N = 2 due to experimental limitations, so that the beam spot is $\Delta\rho$ = 43μm. After determining the values of $A_n$, $k_{pn}$ e $k_{zn}$, we separate the terms of the sum (superposition) getting five fields $\Psi_n(\rho, \phi, z)$ with n = 0, 1, ..., 5.

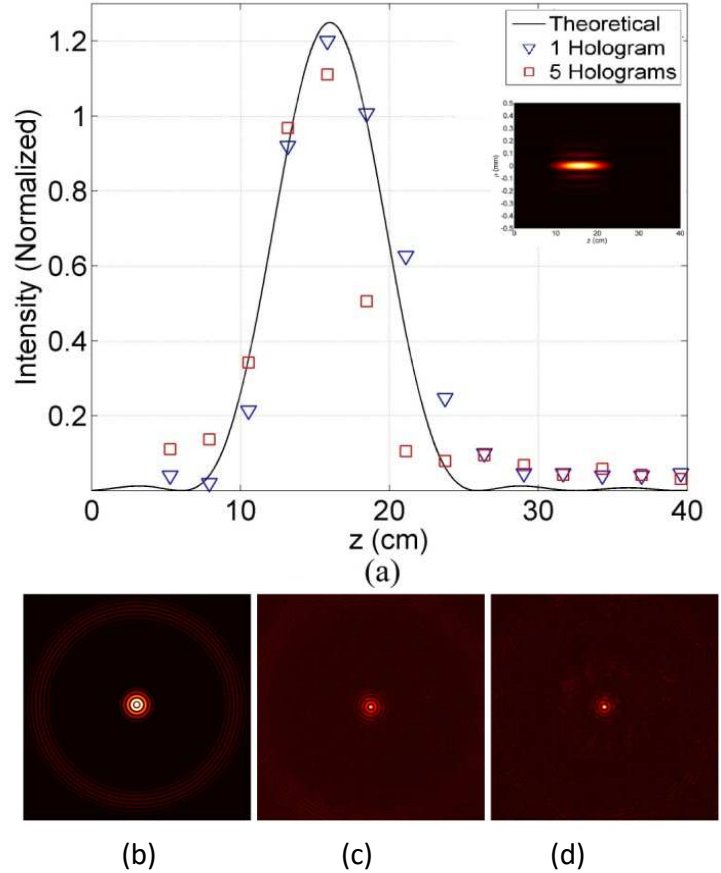

Fig. 8. (a) Longitudinal intensity pattern for the FW described by Eq. (5) and (11), together with its corresponding orthogonal projections; compared to the reconstruction of FW using 1 and 5 holograms and transverse patterns in the maximum intensity position to (b) Eq. (5) (c) one hologram, and (d) 5 holograms.

For each field, we made an individual CGH of the each Bessel beam and we generate its beam (object beam); then using the sequential generation of holograms, described above, let the PRC crystal (holographic recording medium) be exposed to the pattern of each object beam and reference beam by a time interval given by the graph of Figure 6. This time, the beams are superimposed resulting in sequential photorefractive holograms. We considered the following exposure times of each beam: $t_1$=19s, $t_2$=11,5s, $t_3$=9s, $t_4$=6,5s, $t_5$=4s. After recording the five holograms, the object beam was blocked and only the reference beam focused on crystal and affected the simultaneous reading of the holograms, hence the diffracted beam reconstructs the FW expected by the superposition of these Bessel beams. The results are shown in Figure 8.

Similarly, we present another case to a function of type:

$$F(z) = \begin{cases} 1, & l_1 \leq z \leq l_2 \\ 1, & l_3 \leq z \leq l_4 \\ 0, & elsewhere \end{cases} \quad (12)$$

where $l_1$ = 5cm, $l_2$ = 16cm, $l_3$ = 26cm e $l_4$ = 37cm and L = 40cm. And we obtained the experimental generation FW, as shown in Figure 9, with an excellent result compared to theoretical description.

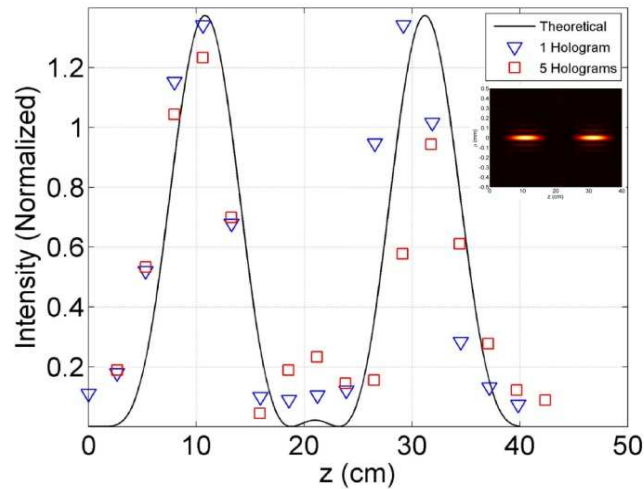

(a)

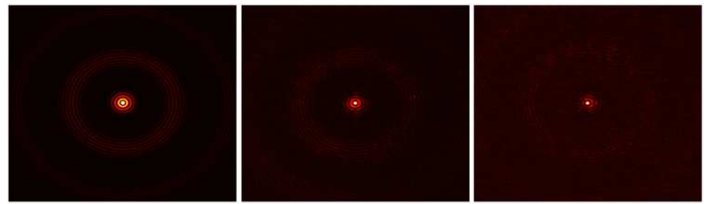

(b)  (c)  (d)

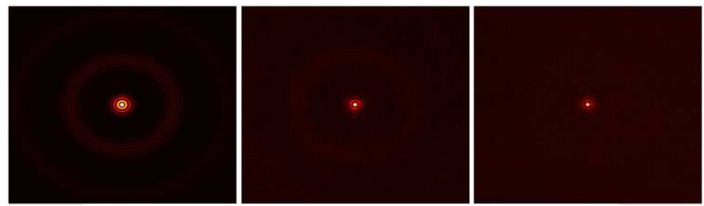

(e)  (f)  (g)

Fig. 9. (a) Longitudinal intensity pattern for the FW described by Eq. (5) and (12), together with its corresponding orthogonal projections; compared to the reconstruction of FW using 1 and 5 holograms and transverse patterns in the maximum intensity position to (b) and (e), Eq. (5); (c) and (f), one hologram; (d) and (g), 5 holograms..

## 4. CONCLUSIONS

The use of photorefractive holography to generate non-diffracting beams was effective, as presented in our results, and shows promise with generating arrays of non-diffracting beams and superposition of nondiffracting beams, where the recording and reconstruction of beams are exclusively optical. We believe the most interesting aspect of this work was the use of photorefractive holography to generate Frozen Waves.

Equally important was the high resolution and high speed record presented by these holographic recording media. A BSO crystal property of great value has fast response time for holographic grating formation (ms), however, it also has a high recombination rate that erases the holographic grating quickly. One advantage of generating non-diffracting beams using the PRC regarding the use of SLM is the direct generation of the beam without the need of 4-f system, besides having far superior to any resolution SLM display, its limitation is only due standard light resolution used to record the hologram, in this case, the diffraction limit of light.

**Funding.** This research is supported by the UFABC, CAPES, FAPESP (grant 09/11429-2) and CNPQ (grants 476805/2012-0 and 313153/2014-0).

**Acknowledgment**. We thank the Mikiya Muramatsu for many stimulating discussions.